\documentclass[journal]{IEEEtran}
\usepackage[dvipdfmx]{graphicx}
\usepackage{amsmath}
\usepackage{amssymb}
\usepackage{theorem}

\theorembodyfont{\rmfamily}
\newtheorem{theorem}{Theorem}
\newtheorem{lemma}[theorem]{Lemma}

\newtheorem{corollary}[theorem]{Corollary}
\newtheorem{remark}[theorem]{Remark}
\newtheorem{proposition}[theorem]{Proposition}
\newcommand{\qed}{\hfill$\square$}

\newcommand{\markov}{\leftrightarrow}
\newcommand{\bol}[1]{\mathbf{#1}}
\newcommand{\rom}[1]{\mathrm{#1}}

\begin{document}

\title{Secret Key Agreement from Vector Gaussian
Sources by Rate Limited Public Communication\thanks{A part of this paper was presented at 2010 IEEE International
Symposium on Information Theory in Austin U.S.A..}}

\author{Shun~Watanabe,~\IEEEmembership{Member,~IEEE,}
and Yasutada Oohama ~\IEEEmembership{Member,~IEEE,}         
\thanks{The authors are with the Department
of Information Science and Intelligent Systems, 
University of Tokushima,
2-1, Minami-josanjima, Tokushima,
770-8506, Japan, 
e-mail:\{shun-wata,oohama\}@is.tokushima-u.ac.jp.}

\thanks{Manuscript received ; revised }}

% The paper headers
\markboth{Journal of \LaTeX\ Class Files,~Vol.~6, No.~1, January~2007}%
{Shell \MakeLowercase{\textit{et al.}}: Bare Demo of IEEEtran.cls for Journals}

\maketitle
\begin{abstract}
%\boldmath
We investigate the secret key agreement
from correlated vector Gaussian sources in which the legitimate
parties can use the public communication with limited rate.
For the class of protocols with the one-way public communication,
we show that the optimal trade-off between
the rate of key generation and the rate of the public 
communication is characterized as an optimization problem
of a Gaussian random variable.
The characterization is derived by using the
enhancement technique introduced by
Weingarten {\em et.~al.}
for MIMO Gaussian broadcast channel.
\end{abstract}

\begin{IEEEkeywords}
Enhancement Technique, Entropy Power Inequality, Extremal Inequality,
Key Agreement, Rate Limited Public Communication
Privacy Amplification, Vector Gaussian Sources,
\end{IEEEkeywords}

\IEEEpeerreviewmaketitle

%%%%%%%%%%%% Introduction %%%%%%%%%
\section{Introduction}

\begin{figure}
\centering
\includegraphics[width=\linewidth]{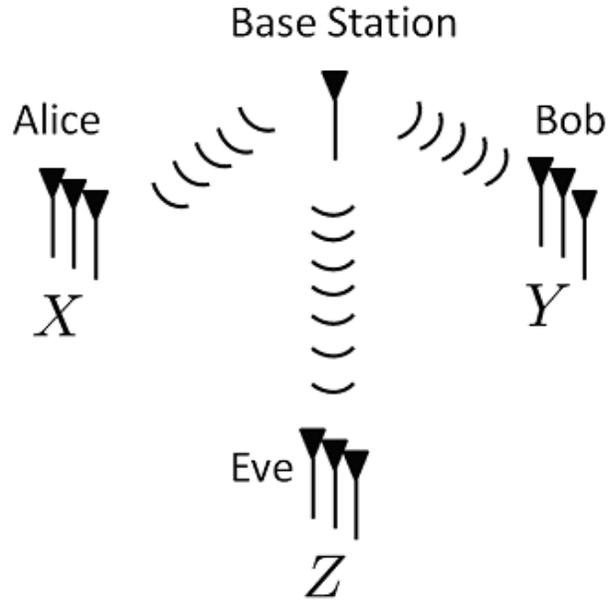}
\caption{An example of the scenarios treated in this paper.
The legitimate parties, Alice and Bob, and an eavesdropper, Eve,
receive (vector) signals from the base station respectively.
Alice and Bob generate a secret key from their received signals
$X$ and $Y$ by using the
public communication.}
\label{Fig:scenario}
\end{figure}

Key agreement is one of the most important 
problems in the cryptography, and it has been
extensively studied in the information theory 
for discrete sources (e.g.~\cite{ahlswede:93,csiszar:00,csiszar:04})
since the problem formulation
by Maurer \cite{maurer:93}.
Recently, the confidential message 
transmission \cite{wyner:75,csiszar:78} in 
the MIMO wireless communication has attracted 
considerable attention  as a
practical problem setting 
(e.g.~\cite{liang:08,liu:09,bustin:09,ly:09,liu:09b,ekrem:09,ekrem:10,liu:10,khisti:10}). 
Although the key agreement in the context of the
wireless communication has also attracted
considerable attention recently \cite{bloch:08},
the key agreement from analog sources has not
been studied sufficiently compared to
the confidential message transmission.
As a fundamental case of the key agreement
from analog sources, we consider the key
agreement from correlated vector Gaussian
sources in this paper. More specifically,
we consider the problem in which the legitimate 
parties, Alice and Bob, and an eavesdropper, Eve,
have correlated vector Gaussian sources respectively,
and Alice and Bob share a secret key from their sources
by using the public communication.
Recently, the key agreement from Gaussian sources
has attracted considerable attention in the context
of the quantum key distribution \cite{grosshans:03},
which is also a motivation to investigate
the present problem.
Fig.~\ref{Fig:scenario} illustrates a scenario
we are considering.
%%%%%%%% Fig %%%%%%%%%%%%%%%

Typically, the first step of the key agreement protocol
from analog sources is the quantization of the sources.
In literatures (e.g.~see \cite{bloch:08,assche:04,assche:book}), 
the authors used the scalar quantizer, 
i.e., the observed source is quantized in each time instant.
Using the finer quantization, we can expect the higher key
rate in the protocol, where the key rate is the ratio between the length of
the shared key and the block length of
the sources that are used in the protocol.
However, there is a problem such that the finer quantization
might increase the rate of the public communication in the protocol.
Although the public communication is usually regarded as a cheap resource
in the context of the key agreement problem, 
it is limited by a certain amount in practice.
Therefore, we consider the key agreement protocols with 
the rate limited public communication in this paper.
The purpose of this paper is to clarify the trade-off
between the key rate and the public communication rate
of the key agreement protocol from vector
Gaussian sources.

The key agreement by rate limited public communication
was first considered by Csisz\'ar and 
Narayan for discrete sources \cite{csiszar:00}.
For the class of protocols with one-way public communication,
they characterized the optimal trade-off between the key rate and
the public communication rate in terms of the information
theoretic quantities, i.e., they derived the so-called
single letter characterization.
However, there are two difficulties to extend their 
result to the vector Gaussian sources.

First, the direct part of the proof in \cite{csiszar:00}
heavily relies on the finiteness of the alphabets
of the sources, and cannot be applied to continuous sources.
This difficulty was solved by the authors in
\cite{watanabe:09c}, and this result will be
also used in this paper.

Second, although
the converse part of Csisz\'ar and Narayan's characterization
can be easily extended to continuous sources,
the characterization is not computable because
the characterization involves auxiliary random variables
and the ranges of those random variables are unbounded
for continuous sources.

In \cite{watanabe:09c} for scalar Gaussian sources,
the authors showed that Gaussian auxiliary random
variables suffice, and derived a closed form expression
of the optimal trade-off.
In the problem for scalar Gaussian sources,
we first solved the problem in which the sources are
degraded, i.e., Alice's source, Bob's source, and
Eve's source form a Markov chain in this order.
Then, we reduced the general case to the degraded
case by using the fact that scalar Gaussian correlated
sources are stochastically degraded \cite{cover}.

In this paper for vector Gaussian sources,
we show that Gaussian auxiliary random
variables suffice, and characterize the
optimal trade-off in terms of the
(covariance) matrix optimization problem.
One of difficulties to show our result is that
vector Gaussian sources are not
stochastically degraded in general, and cannot
be reduced to the degraded case in the same manner
as scalar Gaussian sources.
To circumvent this difficulty, we utilize the 
enhancement technique introduced by
Weingarten {\em et al}.~\cite{weingarten:06}.

%There is another difficulty to show our result.
%In vector Gaussian multiterminal problems, we usually
%need to upper bound differences of 
%(conditional) differential entropies,
%which are known as extremal inequalities 
%\cite{liu:07,weingarten:09}.
%In our problem, 
%we need to show an extremal inequality
%which has not appeared in the
%literatures \cite{liu:07,weingarten:09}.
%We circumvent this difficulty by using
%a change of variable and by reducing to
%more tractable form of an extremal inequality,
%which has appeared in the literature
%(see Remark \ref{remark:extremal}).
%The authors believe that this bounding technique
%itself is also interesting contribution.

The rest of the paper is organized as follows:
In Section \ref{sec:preliminaries}, we explain
our problem formulation.
In Section \ref{sec:main}, we show
our main results and some numerical examples.
In Sections \ref{sec:aligned-case} and
\ref{sec:proof-general}, our main results
are proved.
Finally, in Section \ref{sec:conclusion},
the conclusion and the future research agenda
are discussed.

%%%%%%%%%%%%%%%% Problem Formulation %%%%%%%%%%%%%%%
\section{Problem Formulation}
\label{sec:preliminaries}

Let $X$, $Y$, and $Z$ be correlated 
vector Gaussian sources on $\mathbb{R}^{m_x}$,
$\mathbb{R}^{m_y}$, and $\mathbb{R}^{m_z}$
respectively,
where $\mathbb{R}$ is the set of real
numbers.
Then, let $X^n$, $Y^n$, and $Z^n$ be i.i.d.
copies of $X$, $Y$, and $Z$ respectively.
Throughout the paper, upper case letters indicate
random variables, and the corresponding lower case
letters indicate their realizations.
We also use the following notations throughout the paper:
$\Sigma$ designates the covariance matrix of $(X,Y,Z)$.
$\Sigma_x$, $\Sigma_{xy}$, and $\Sigma_{y|x}$ designate
$\mathbb{E}[X^TX]$, $\mathbb{E}[X^T Y]$, and 
the conditional covariance of $Y$ given $X$ etc..
$N \sim {\cal N}(0, A)$ means that the random variable
$N$ is a Gaussian vector with zero mean and covariance
matrix $A$.  We use $|A|$ to denote the determinant
of the matrix $A$,
$\left|\frac{A}{B}\right|$ to denote 
$\frac{|A|}{|B|}$, and we denote
$A \preceq B$ ($A \prec B$) if the matrix $B-A$ is
positive semidefinite (definite). 
Throughout the paper, we assume that $\Sigma \succ 0$.

\begin{figure}
\centering
\includegraphics[width=\linewidth]{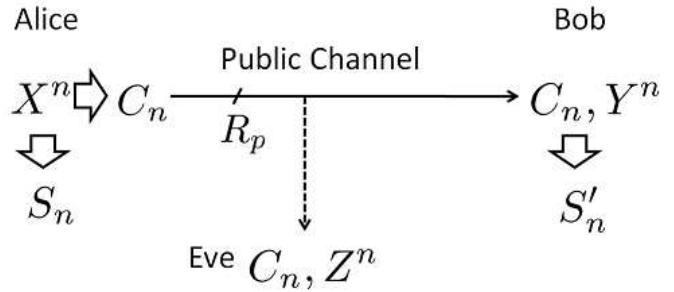}
\caption{The key agreement protocol with one-way public communication.
First, Alice sends the message $C_n$ to Bob over the public channel, which 
might be eavesdropped by Eve. Then, Alice compute the key
$S_n$ and Bob compute the key $S_n^\prime$. The rate of the public
communication is limited by $R_p$.}
\label{Fig:protocol}
\end{figure}

Although Alice and Bob can use public communication
interactively in general, we concentrate on the
class of key agreement protocols in which only Alice 
sends a message to Bob over the public channel.
First, Alice computes the message
$C_n$ from $X^n$ and sends the
message to Bob
over the public channel.
Then, she also compute the key $S_n$.
Bob compute the key $S_n^\prime$
from $Y^n$ and $C_n$.
Fig.~\ref{Fig:protocol} illustrates the protocol
with one-way public communication.
%%%%%%%% Fig %%%%%%%%%%%%%%%

The error probability of the protocol
is defined by 
\begin{eqnarray*}
\varepsilon_n := \Pr\{ S_n \neq S_n^\prime \}.
\end{eqnarray*}
The security of the protocol is measured by
the quantity
\begin{eqnarray*}
\nu_n := \log|{\cal S}_n| - 
 H(S_n|C_n,Z^n),
\end{eqnarray*}
where ${\cal S}_n$ is the range of
the key $S_n$.

In this paper, we are interested in the trade-off
between the public communication rate $R_p$
and the key rate $R_k$.
The rate pair $(R_p,R_k)$ is defined to 
be achievable if there exists a sequence of
protocols satisfying
\begin{eqnarray*}
\lim_{n \to \infty} \varepsilon_n &=& 0, \\
\lim_{n \to \infty} \nu_n &=& 0, \\
\limsup_{n \to \infty} \frac{1}{n} \log|{\cal C}_n| &\le& R_p,\\
\liminf_{n \to \infty} \frac{1}{n} \log|{\cal S}_n| &\ge& R_k,
\end{eqnarray*}
where ${\cal C}_n$ is the range of the message $C_n$
transmitted over the public channel.
Then, the achievable rate region is defined as
\begin{eqnarray*}
{\cal R}(X,Y,Z) :=
\{(R_p,R_k) : \mbox{$(R_p,R_k)$ is achievable} \}.
\end{eqnarray*}

In \cite{watanabe:09c}, the authors showed a closed 
form expression of ${\cal R}(X,Y,Z)$ for the scalar problem, i.e.,
$m_x = m_y = m_z = 1$. In the next section, we show that
the achievable rate region for the vector problem can be characterized as a
(covariance) matrix optimization problem.

%%%%%%%%%%%%%% Main Result %%%%%%%%%%%%%%%%%%%%%%%5
\section{Main Result}
\label{sec:main}

\subsection{Main Theorems}

In this section, we show our main results.
Since the security quantities $\varepsilon_n$ and 
$\nu_n$ only depend on the marginal distributions of
$(X,Y)$ and $(X,Z)$ respectively, it suffice to consider
$(X,Y,Z)$ of the form
\begin{eqnarray*}
Y &=& B X + W_y, \\
Z &=& E X + W_z,
\end{eqnarray*}
where $B \in \mathbb{R}^{m_y \times m_x}$,
$E \in \mathbb{R}^{m_z \times m_x}$, 
$W_y \sim {\cal N}(0, I_{m_y})$ and
$W_z \sim {\cal N}(0, I_{m_z})$.
In the rest of this paper, we omit the subscript
of the identity matrix if the dimension is obvious
from the context. 

One of the main results of this paper is 
the following.

\begin{theorem}
\label{theorem:general-case}
Let ${\cal R}_G(X,Y,Z)$ be the set
of all rate pairs $(R_p,R_k)$ satisfying
\begin{eqnarray*}
R_p
&\ge& \frac{1}{2}\log \left|\frac{\Sigma_x }{\Sigma_{x|u} }\right|
 - \frac{1}{2}\log \left|\frac{B \Sigma_x B^T + I}{
 B \Sigma_{x|u} B^T + I}\right|, \\
R_k
&\le& \frac{1}{2}\log \left|\frac{ B \Sigma_x B^T + I}{
 B \Sigma_{x|u} B^T + I}\right|
- \frac{1}{2}\log \left|\frac{ E \Sigma_x E^T + I}{
 E \Sigma_{x|u} E^T + I}\right|
\end{eqnarray*}
for some $0 \prec \Sigma_{x|u} \preceq \Sigma_x$.
Then, we have
\begin{eqnarray*}
{\cal R}(X,Y,Z) = {\cal R}_G(X,Y,Z).
\end{eqnarray*}
\end{theorem}

We are also interested in the asymptotic behavior
of the function
\begin{eqnarray}
R_k(R_p) := \sup\{ R_k:~(R_p,R_k) \in {\cal R}(X,Y,Z) \}.
\label{asymptotic-bahavior}
\end{eqnarray}
Following the approach in \cite{liu:09b}, we can obtain
a closed form expression of 
$\lim_{R_p \to \infty}R_k(R_p)$ as follows.
Let $\phi_i,i=1,\ldots,m_x$ be the generalized
eigenvalues \cite[Chapter 6.3]{strang} of the matrices
\begin{eqnarray*}
\left(
\Sigma_x^{\frac{1}{2}} B^T B \Sigma_x^{\frac{1}{2}} + I_{m_x},
\Sigma_x^{\frac{1}{2}} E^T E \Sigma_x^{\frac{1}{2}} + I_{m_x}
\right).
\end{eqnarray*}
Without loss of generality, we may assume that
these generalized eigenvalues are ordered as
\begin{eqnarray}
\label{eq:order-of-eigenvalues}
\phi_1 \ge \cdots \ge \phi_\rho > 1 
\ge \phi_{\rho+1} \ge \cdots \ge \phi_{m_x},
\end{eqnarray} 
i.e., a total of $\rho$ of them are assumed to be greater than $1$.
Then, we have
\begin{eqnarray}
\lim_{R_p \to \infty}R_k(R_p)  
&=& \max_{0 \preceq \Sigma_{x|u} \preceq \Sigma_x}
 \left[
 \frac{1}{2}\log \left|\frac{ B \Sigma_x B^T + I}{
 B \Sigma_{x|u} B^T + I}\right| \right. \nonumber \\
&& \left. ~~~~~~~~~~~~~- \frac{1}{2}\log \left|\frac{ E \Sigma_x E^T + I}{
 E \Sigma_{x|u} E^T + I}\right|
 \right] \nonumber \\
&=& \frac{1}{2}\sum_{i =1}^\rho \log \phi_i.
\label{eq:closed-form-upper}
\end{eqnarray}
Since Eq.~(\ref{eq:closed-form-upper}) can be proved
almost in the same manner as \cite[Theorem 3]{liu:09b},
we omit a proof.

When $m_x=m_y=m_z$ and both $B$ and $E$
are invertible, it suffice to consider 
the case in which
\begin{eqnarray}
\label{eq:aligned-y}
Y &=& X + W_y, \\
\label{eq:aligned-z}
Z &=& X + W_z,
\end{eqnarray}
where the covariance matrices
$\Sigma_{W_y}$ and $\Sigma_{W_z}$ are
not necessarily identity but are invertible.
Following \cite{weingarten:06}, we call this
case the aligned case.
As is usual with the vector Gaussian 
problems (e.g.~\cite{weingarten:06}),
the general statement (Theorem \ref{theorem:general-case})
is shown by detouring the statement for
the aligned case.

\begin{theorem}
\label{theorem:aligned-case}
Let 
${\cal R}^*_G(X,Y,Z)$ be the set
of all rate pairs $(R_p,R_k)$ satisfying
\begin{eqnarray*}
R_p &\ge&
I_p(\Sigma_{x|u}) \\
&:=& \frac{1}{2}\log \left|\frac{\Sigma_x }{\Sigma_{x|u} }\right|
 - \frac{1}{2}\log \left|\frac{\Sigma_x  + \Sigma_{W_y}}{
 \Sigma_{x|u} + \Sigma_{W_y}}\right|, \\
R_k &\le&
I_k(\Sigma_{x|u}) \\
&:=& \frac{1}{2}\log \left|\frac{ \Sigma_x + \Sigma_{W_y}}{
 \Sigma_{x|u}  + \Sigma_{W_y}}\right|
- \frac{1}{2}\log \left|\frac{ \Sigma_x  + \Sigma_{W_z}}{
 \Sigma_{x|u} + \Sigma_{W_z}}\right|
\end{eqnarray*}
for some $0 \prec \Sigma_{x|u} \preceq \Sigma_x$.
Then, we have
\begin{eqnarray*}
{\cal R}(X,Y,Z) = {\cal R}^*_G(X,Y,Z).
\end{eqnarray*}
\end{theorem}

Theorem \ref{theorem:aligned-case} is shown 
in Section \ref{sec:aligned-case}
and Theorem \ref{theorem:general-case} 
is shown in Section \ref{sec:proof-general}
by using Theorem \ref{theorem:aligned-case}.

%%%%%%%%%%%%%%%% Numerical %%%%%%%%%%%%%%%%%%%%%%%%
\subsection{Numerical Examples}

In this section, we show some numerical
example to illustrate Theorem \ref{theorem:general-case}.
In general,
calculation of ${\cal R}_G(X,Y,Z)$ involves a
nonconvex optimization problem and is not tractable.
However for $m_x \ge 2$ and $m_y=m_z=1$, 
following the method in \cite{weingarten:06b} 
(see also \cite{ly:09}), we can transform the calculation of
${\cal R}_G(X,Y,Z)$ into tractable form.

For $m_x \ge 2$ and $m_y=m_z=1$, we have
\begin{eqnarray*}
I_p(\Sigma_{x|u}) &=& \frac{1}{2} \log \left|
 \frac{\Sigma_x}{\Sigma_{x|u}} \right|
  - \frac{1}{2} \log \frac{b \Sigma_x b^T +1}{b \Sigma_{x|u} b^T + 1},\\
I_k(\Sigma_{x|u}) &=& \frac{1}{2} \log
 \frac{b \Sigma_x b^T + 1}{b \Sigma_{x|u} b^T + 1}
 - \frac{1}{2} \log \frac{e \Sigma_x e^T + 1}{e \Sigma_{x|u} e^T + 1},
\end{eqnarray*}
where $b, e \in \mathbb{R}^{m_x}$.
Noting the relation
\begin{eqnarray*}
\frac{e \Sigma_{x|u} e^T + 1}{b \Sigma_{x|u} b^T + 1}
 = 1 + \frac{e \Sigma_{x|u} e^T - b \Sigma_{x|u} b^T + 1}{ b \Sigma_{x|u} b^T + 1},
\end{eqnarray*}
we set
\begin{eqnarray*}
s &=& b \Sigma_{x|u} b^T, \\
t &=& \frac{e \Sigma_{x|u} e^T - b \Sigma_{x|u} b^T + 1}{ b \Sigma_{x|u} b^T + 1}.
\end{eqnarray*}
Let 
\begin{eqnarray*}
I_p(\Sigma_{x|u},s) &:=& \frac{1}{2} \log \left| \frac{\Sigma_x}{\Sigma_{x|u}}\right|
  - \frac{1}{2} \log (b \Sigma_x b^T + 1) \\
&&~~~~~~~~~~~~~~~~ + \frac{1}{2} \log(1+s), \\
I_k(t) &=& \frac{1}{2}\log \frac{b\Sigma_x b^T + 1}{e 
 \Sigma_x e^T + 1} + \frac{1}{2}\log (1 + t).
\end{eqnarray*}
Then we can easily find that
\begin{eqnarray*}
\lefteqn{
\hspace{-35mm} {\cal R}_G(X,Y,Z) } \\
= \{ (R_p,R_k):~R_p &\ge& I_p(\Sigma_{x|u},s), \\
 R_k &\le&  I_k(t), \\
0 \prec \Sigma_{x|u} &\preceq& \Sigma_x, \\
t(b \Sigma_{x|u}b^T + 1) &\le& e\Sigma_{x|u}e^T - b \Sigma_{x|u} b^T ,\\
b \Sigma_{x|u} b^T &\le& s \}.
\end{eqnarray*}
For fixed $(s,t)$, the optimization problem
\begin{eqnarray*}
\mbox{minimize} && I_p(\Sigma_{x|u},s) \\
\mbox{subject to} && t(b \Sigma_{x|u} b^T +1) 
 \le e \Sigma_{x|u} e^T - b \Sigma_{x|u} b^T \\
&& b \Sigma_{x|u} b^T \le s \\
&& 0 \prec \Sigma_{x|u} \preceq \Sigma_x
\end{eqnarray*}
is a convex problem.
By sweeping $(s,t)$, we can calculate
the region ${\cal R}_G(X,Y,Z)$.

\begin{figure}
\centering
\includegraphics[width=\linewidth]{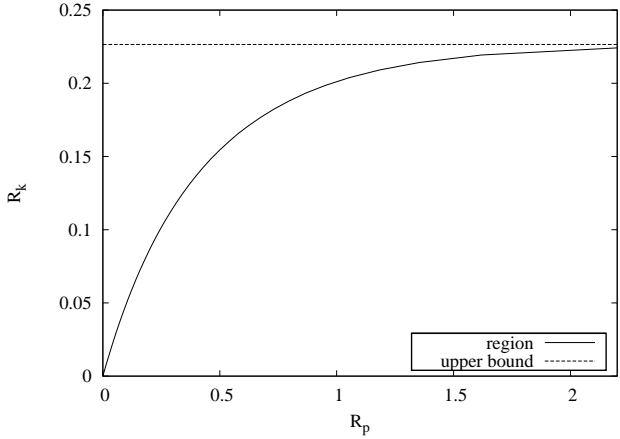}
\caption{''region`` is 
$R_k(R_p)$ defined in Eq.~(\ref{asymptotic-bahavior}) for the sources given
by Eq.~(\ref{eq:example-1}).
''upper bound'' is the quantity $I(X;Y) - I(X;Z)$.
}
\label{Fig:degraded}
\end{figure}

For 
\begin{eqnarray}
\label{eq:example-1}
\Sigma_x = \left[
\begin{array}{cc}
2 & 0 \\ 0 & 2
\end{array} \right],
~ b = \left[
\begin{array}{cc}
1 & 0.5
\end{array} \right],
~ e = \left[
\begin{array}{cc}
0.7 & 0.35
\end{array} \right],
\end{eqnarray}
the region ${\cal R}_G(X,Y,Z)$ is plotted 
in Fig.~\ref{Fig:degraded}.
Note that this case is degraded in the 
sense of \cite[Definition 1]{weingarten:09},
i.e., $X \markov Y \markov Z$ by appropriately
choosing the correlation between $(Y,Z)$.
In this case, the function $R_k(R_p)$
converges to $I(X;Y) - I(X;Z)$ as 
$R_p$ increases.
%%%%%%%%%%%% Fig 1 %%%%%%% %%%%%%%%%%%%%%%%
%%%%%%%%%%%%%%%%%%%%%%%%%%%%%%%%%%%%%%%%%%

\begin{figure}
\centering
\includegraphics[width=\linewidth]{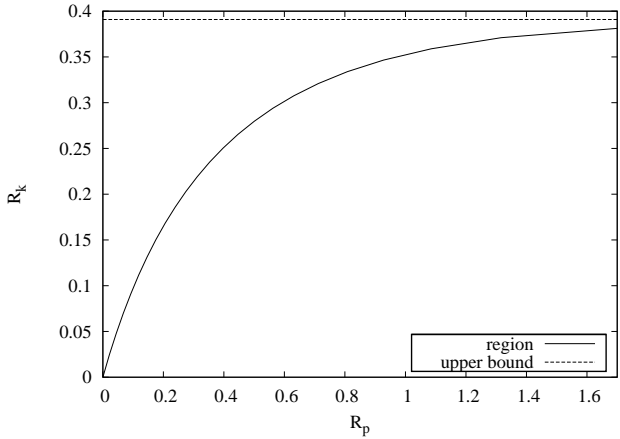}
\caption{''region`` is 
$R_k(R_p)$ defined in Eq.~(\ref{asymptotic-bahavior}) for the sources given
by Eq.~(\ref{eq:example-2}). ``upper bound
is $\lim_{R_p \to \infty} R_k(R_p)$ which is
explicitly given by Eq.~(\ref{eq:closed-form-upper}).
}
\label{Fig:non_degraded}
\end{figure}

For 
\begin{eqnarray}
\label{eq:example-2}
\Sigma_x = \left[
\begin{array}{cc}
2 & 0 \\ 0 & 2
\end{array} \right],
~ b = \left[
\begin{array}{cc}
1 & 0.5
\end{array} \right],
~ e = \left[
\begin{array}{cc}
0.5 & 1
\end{array} \right],
\end{eqnarray}
the region ${\cal R}_G(X,Y,Z)$ is plotted 
in Fig.~\ref{Fig:non_degraded}.
Note that $(X,Y,Z)$ in this example is not degraded.
Although $I(X;Y) - I(X;Z) = 0$ in this example, 
Fig.~\ref{Fig:non_degraded} clarifies that appropriate
quantization enables Alice and Bob to share a secret
key at positive key rate.    
%%%%%%%%%%%% Fig 2 %%%%%%%%%%%%%%%%%%%%%%%
%%%%%%%%%%%%%%%%%%%%%%%%%%%%%%%%%%%%%%%%%%

For non-degraded case, 
$R_k(R_p)$ converges to the quantity
given by Eq.~(\ref{eq:closed-form-upper}) instead of
$I(X;Y) - I(X;Z)$ as $R_p$ increases,
and it is also plotted in Fig.~\ref{Fig:non_degraded}.

%%%%%%%%%%%%%%%%%%%%%% Aligned Case %%%%%%%%%%%%%%%%55

\section{Proof of Theorem \ref{theorem:aligned-case}}
\label{sec:aligned-case}

\subsection{Direct Part}
\label{subsec:direct-aligned}

In \cite{watanabe:09c}, the present authors proved
the following proposition, which is an extension
of \cite[Theorem 2.6]{csiszar:00} to continuous sources.

\begin{proposition}
\label{proposition:direct-part}
For an auxiliary random variable $U$ satisfying the Markov chain
\begin{eqnarray*}
U \markov X \markov (Y,Z),
\end{eqnarray*}
let $(R_p,R_k)$ be a rate pair such that
\begin{eqnarray*}
R_p &\ge& I(U;X) - I(U;Y), \\
R_k &\le& I(U;Y) - I(U;Z).
\end{eqnarray*}
Then, we have $(R_p,R_k) \in {\cal R}(X,Y,Z)$.
\end{proposition}

The direct part of Theorem \ref{theorem:aligned-case} is shown by
taking Gaussian auxiliary random variable $U$ 
such that the conditional covariance matrix
of $X$ given $U$ is $\Sigma_{x|u}$
in Proposition \ref{proposition:direct-part}.
\qed

%%%%%% Converse %%%%%%%%%%%%%%
\subsection{Converse Part}

In the converse proof, we will use the following
Proposition and Corollary. 
The proposition was shown for discrete
sources in \cite[Theorem 2.6]{csiszar:00},
and it can be shown almost in the same manner
for continuous sources.
\begin{proposition}
(\cite{csiszar:00}) 
\label{prop:cn-bound}
Suppose that a 
rate pair $(R_p,R_k)$ is included in
${\cal R}(X,Y,Z)$. Then, 
there exist
auxiliary random variables $U$ and $V$ satisfying
\begin{eqnarray}
\label{public-lower}
R_p &\ge& I(U;X|Y), \\
R_k &\le& I(U;Y|V) - I(U;Z|V),
\end{eqnarray}
and the Markov chain
\begin{eqnarray}
\label{markov-condition}
V \markov U \markov X \markov (Y,Z).
\end{eqnarray}
\end{proposition}

For degraded sources, we can simplify 
the above proposition
(see \cite[Appendix B]{watanabe:09c} for a proof).
\begin{corollary}
\label{coro:cn-bound}
Suppose that $(X,Y,Z)$ is degraded, i.e., $X \markov Y \markov Z$.
If $(R_p,R_k) \in {\cal R}(X,Y,Z)$, then
there exists an auxiliary random variable $U$ satisfying
\begin{eqnarray}
\label{eq:coro-public-upper}
R_p &\ge& I(U;X|Y)
= I(U;X) - I(U;Y), \\
\label{eq:coro-key-upper}
R_k &\le& I(U;Y|Z)
 = I(U;Y) - I(U;Z),
\end{eqnarray}
and the Markov chain
\begin{eqnarray}
\label{eq:markov-2}
U \markov X \markov Y \markov Z.
\end{eqnarray}
\end{corollary}

We show a converse proof of
Theorem \ref{theorem:aligned-case}
by contradiction.
Suppose that there exists a rate pair such that
$(R_p^o,R_k^o) \in {\cal R}(X,Y,Z)$
and $(R_p^o,R_k^o) \notin {\cal R}_G^*(X,Y,Z)$,
where we assume $R_k^o > 0$ to avoid
the trivial case.
Then, there exists $0 \prec \Sigma_{x|u}^o \preceq \Sigma_x$
such that $I_p(\Sigma_{x|u}^o) \le R_p^o$.
Therefore, we can write 
\begin{eqnarray}
\label{eq:key-rate-equality}
R_k^o = R_k^* + \delta
\end{eqnarray} 
for some $\delta > 0$, where $R_k^*$ is given by
the optimal value of
\begin{eqnarray}
\mbox{maximize} && I_k(\Sigma_{x|u}) \nonumber\\
\mbox{subject to} && I_p(\Sigma_{x|u}) \le R_p^o, 
 \label{eq:opt1} \\
 && 0 \prec \Sigma_{x|u} \preceq \Sigma_x.
 \nonumber
\end{eqnarray}
An optimal solution $\Sigma_{x|u}^*$ of
this optimization problem satisfies
the Karash-Kuhn-Tucker (KKT) 
condition (see Appendix \ref{appendix:kkt} for the
derivation)
\begin{eqnarray}
\mu (\Sigma_{x|u}^*)^{-1}
 &+& (\Sigma_{x|u}^* + \Sigma_{W_z})^{-1}  \nonumber \\
 &=& 
(1+\mu) (\Sigma_{x|u}^* + \Sigma_{W_y})^{-1} + M, \nonumber \\
 \label{eq:kkt-1} \\
\label{eq:kkt-2}
M(\Sigma_x - \Sigma_{x|u}^*) &=& 0, \\
\label{eq:kkt-3}
\mu(R_p^o - I_p(\Sigma_{x|u}^*)) &=& 0,
\end{eqnarray}
where $\mu \ge 0$ and $M \succeq 0$.
From Eqs.~(\ref{eq:key-rate-equality}) 
and (\ref{eq:kkt-3}), we have
\begin{eqnarray}
\label{eq:contradiction}
R_k^o - \mu R_p^o =
 I_k(\Sigma_{x|u}^*) - \mu I_p(\Sigma_{x|u}^*)
 + \delta.
\end{eqnarray}
We shall find a contradiction to Eq.~(\ref{eq:contradiction})
by showing that for any $(R_p,R_k) \in {\cal R}(X,Y,Z)$
\begin{eqnarray}
\label{eq:goal}
R_k - \mu R_p \le I_k(\Sigma_{x|u}^*)
 - \mu I_p(\Sigma_{x|u}^*).
\end{eqnarray}

The proof of Eq.~(\ref{eq:goal}) roughly consists
of three steps: In the first step,
we reduce the proof for the non-degraded sources
to that for the degraded sources by using the
enhancement technique introduced by
Weingarten {\em et.~al.} \cite{weingarten:06}.
In the second step, we change the variable so that
we can use the entropy power inequality (EPI).
In the last step, we derive an upper bound
on $R_k - \mu R_p$ by using the EPI,
which turn out to be tight.

\noindent{\em Step 1:}
In this step, in order to reduce the 
proof for the non-degraded sources to
that for the degraded sources, 
we introduce the covariance matrix
$\Sigma_{\tilde{W}_y}$ satisfying
\begin{eqnarray}
\lefteqn{ (1+\mu) (\Sigma_{x|u}^* + \Sigma_{\tilde{W}_y})^{-1} } \nonumber \\
\label{eq:enhancement}
 &=& (1+\mu) (\Sigma_{x|u}^* + \Sigma_{W_y})^{-1} +  M.
\end{eqnarray}
Then, we have (see Appendix \ref{appendix:enhanced-1-2} for a proof)
\begin{eqnarray}
\label{eq:enhanced-1}
0 \prec \Sigma_{\tilde{W}_y} \preceq \Sigma_{W_y},&&\\
\label{eq:enhanced-2}
\Sigma_{\tilde{W}_y} \preceq \Sigma_{W_z}. &&
\end{eqnarray}
Let $\tilde{W}_y$ be the Gaussian random vector whose 
covariance matrix is $\Sigma_{\tilde{W}_y}$, and let 
\begin{eqnarray}
\tilde{Y} = X + \tilde{W}_y.
\label{eq:enhanced-y}
\end{eqnarray}
From Eq.~(\ref{eq:enhanced-2}), we can find that
the sources $(X,\tilde{Y},Z)$
satisfy $X \markov \tilde{Y} \markov Z$.
Furthermore, from Eq.~(\ref{eq:enhanced-1}), we can also 
find that $X \markov \tilde{Y} \markov Y$, which implies
\begin{eqnarray*}
{\cal R}(X,Y,Z) \subset {\cal R}(X,\tilde{Y},Z).
\end{eqnarray*}
Thus, it suffice to show that 
Eq.~(\ref{eq:goal}) holds for any
$(R_p,R_k) \in {\cal R}(X,\tilde{Y},Z)$.
In steps 2 and 3, we will show that
\begin{eqnarray}
\label{eq:intermidiate}
R_k - \mu R_p \le 
\tilde{I}_k(\Sigma_{x|u}^*) 
  - \mu \tilde{I}_p(\Sigma_{x|u}^*)
\end{eqnarray}
for any $(R_p,R_k) \in {\cal R}(X,\tilde{Y},Z)$,
where 
\begin{eqnarray*}
\tilde{I}_p(\Sigma_{x|u})
&:=& \frac{1}{2}\log \left|\frac{\Sigma_x}{\Sigma_{x|u}}\right|
 - \frac{1}{2}\log \left|\frac{\Sigma_x + \Sigma_{\tilde{W}_y}}{
 \Sigma_{x|u} + \Sigma_{\tilde{W}_y}}\right|, \\
\tilde{I}_k(\Sigma_{x|u})
&:=& \frac{1}{2}\log \left|\frac{\Sigma_x + \Sigma_{\tilde{W}_y}}{
 \Sigma_{x|u} + \Sigma_{\tilde{W}_y}}\right|
- \frac{1}{2}\log \left|\frac{\Sigma_x + \Sigma_{W_z}}{
 \Sigma_{x|u} + \Sigma_{W_z}}\right|.
\end{eqnarray*}
Then, by using the relation
(see Appendix \ref{appendix:preservation} for a proof)
\begin{eqnarray}
\lefteqn{ (\Sigma_x + \Sigma_{\tilde{W}_y})
 (\Sigma_{x|u}^* + \Sigma_{\tilde{W}_y})^{-1} } \nonumber \\
\label{eq:reduction}
&=& (\Sigma_x + \Sigma_{W_y})
 (\Sigma_{x|u}^* + \Sigma_{W_y})^{-1},
\end{eqnarray}
we have $I_k(\Sigma_{x|u}^*) = \tilde{I}_k(\Sigma_{x|u}^*)$
and $I_p(\Sigma_{x|u}^*) = I_p(\Sigma_{x|u}^*)$. Thus, 
Eq.~(\ref{eq:intermidiate}) implies 
that Eq.~(\ref{eq:goal}) holds for any
$(R_p,R_k) \in {\cal R}(X,\tilde{Y},Z)$.

\noindent{\em Step 2:}
First, we show Eq.~(\ref{eq:intermidiate})
for $\mu = 0$. In this case, from
Eqs.~(\ref{eq:kkt-1}) 
and (\ref{eq:enhancement}), we have
$\Sigma_{\tilde{W}_y} = \Sigma_{W_z}$.
Thus, from Corollary \ref{coro:cn-bound}, we have
\begin{eqnarray*}
R_k - 0 \cdot R_p \le
 I(U;Y) - I(U;Z) = 0
 = \tilde{I}_k(\Sigma_{x|u}^*).
\end{eqnarray*}
Thus, we have the assertion.

In order to prove Eq.~(\ref{eq:intermidiate})
for $\mu > 0$,
we change the variable as follows.
Since $(X,\tilde{Y},Z)$ is jointly Gaussian, we can write
\begin{eqnarray*}
X &=& K_{xz} Z + N_1,\\
\tilde{Y} &=& K_{\tilde{y}x} X +  
  K_{\tilde{y}z} Z + N_2 
\end{eqnarray*}
for Gaussian random vectors $N_1, N_2$ with covariance
matrices
\begin{eqnarray*}
\Sigma_{N_1} &=& \Sigma_{x|z} 
 := \Sigma_x - K_{xz} \Sigma_{zx}, \\
\Sigma_{N_2} &=& \Sigma_{\tilde{y}|xz}
 := \Sigma_{\tilde{y}} - K_{\tilde{y}x} \Sigma_{x\tilde{y}}
 - K_{\tilde{y}z} \Sigma_{z\tilde{y}},
\end{eqnarray*}
where the coefficients are given by
\begin{eqnarray*}
K_{xz} = \Sigma_{xz} \Sigma_z^{-1}
\end{eqnarray*}
and
\begin{eqnarray}
\left[ \begin{array}{cc}
K_{\tilde{y}z} & K_{\tilde{y}x}
\end{array} \right]
 = \left[ \begin{array}{cc}
\Sigma_{\tilde{y}z} & \Sigma_{\tilde{y}x}
\end{array} \right]
\left[ \begin{array}{cc}
\Sigma_z & \Sigma_{zx} \\
\Sigma_{xz} & \Sigma_x 
\end{array} \right]^{-1}.
\label{eq:coefficients}
\end{eqnarray}
By noting the relations
\begin{eqnarray*}
I(U;X|\tilde{Y}) &=& I(U;X) - I(U;\tilde{Y}),\\
I(U;\tilde{Y}|Z) &=& I(U;\tilde{Y}) - I(U;Z),\\
I(U;X|Z) &=& I(U;X) - I(U;Z), \\
I(U;X|\tilde{Y}) &=& I(U;X|Z) - I(U;\tilde{Y}|Z)
\end{eqnarray*}
for random variables satisfying
$U \markov X \markov \tilde{Y} \markov Z$, we have
\begin{eqnarray}
\lefteqn{ \tilde{I}_k(\Sigma_{x|u}) - 
 \mu \tilde{I}_p(\Sigma_{x|u}) } \nonumber \\
&=& I(U; \tilde{Y}|Z) - \mu I(U;X|\tilde{Y}) \nonumber \\
&=& (1+\mu) I(U;\tilde{Y}|Z) - \mu I(U;X|Z) \nonumber \\
&=& [(1+\mu) h(\tilde{Y}|Z) - \mu h(X|Z)] \nonumber \\
&& + [\mu h(X|U,Z) - (1+\mu) h(\tilde{Y}|U,Z)] \nonumber \\
&=& [(1+\mu) h(\tilde{Y}|Z) - \mu h(X|Z)]
 - \frac{1+\mu}{2}\log|K_{\tilde{y}x} K_{\tilde{y}x}^T| \nonumber \\
&& + \mu[ h(X|U,Z) - \gamma h(X + K_{\tilde{y}x}^{-1} N_2|U,Z)] \nonumber\\
&=& [(1+\mu) h(\tilde{Y}|Z) - \mu h(X|Z)]
 - \frac{1+\mu}{2}\log|K_{\tilde{y}x} K_{\tilde{y}x}^T| \nonumber \\
&& + \mu[ h(X|U,Z) - \gamma h(X + N_3|U,Z)] \nonumber \\
&=& [(1+\mu) h(\tilde{Y}|Z) - \mu h(X|Z)]
 - \frac{1+\mu}{2}\log|K_{\tilde{y}x} K_{\tilde{y}x}^T| \nonumber \\
&+&  \hspace{-3mm} \mu \left[
 \frac{1}{2} \log (2 \pi e)^m |\Sigma_{x|uz}|
  - \frac{\gamma}{2} 
 \log (2 \pi e)^m |\Sigma_{x|uz} + \Sigma_{N_3}| \right],\nonumber \\
\label{eq:variable-changed}
\end{eqnarray}
where we set $\gamma := \frac{1+\mu}{\mu} > 1$ and
$N_3 := K_{\tilde{y}x}^{-1} N_2$.
It should be noted that
\begin{eqnarray}
\label{eq:invertibility}
|K_{\tilde{y}x}| \neq 0
\end{eqnarray}
for $\mu >0$,
which will be proved in 
Appendix \ref{appendix:invertibility}.

For the change of variable 
\begin{eqnarray*}
\phi: \Sigma_{x|u} \mapsto \Sigma_{x|uz}
 = (\Sigma_{x|u}^{-1} + \Sigma_{W_z}^{-1})^{-1}, 
\end{eqnarray*}
let $\Sigma_{x|uz}^* := \phi(\Sigma_{x|u}^*)$.
From Eqs.~(\ref{eq:kkt-1}) and (\ref{eq:enhancement})
and the relation  
\begin{eqnarray*}
\lefteqn{ \tilde{I}_k(\Sigma_{x|u}) - \mu \tilde{I}_p(\Sigma_{x|u}) } \nonumber \\
&=& \frac{\mu}{2}\log (2 \pi e)^m |\Sigma_{x|u}| 
 + \frac{1}{2} \log (2 \pi e)^m |\Sigma_{x|u} + \Sigma_{W_z}| \nonumber \\
 &&~~~~~~ - \frac{(1+\mu)}{2} \log (2 \pi e)^m |\Sigma_{x|u} + \Sigma_{\tilde{W}_y}| 
\nonumber\\
&& + [(1+\mu)h(\tilde{Y}) - h(Z) - \mu h(X)],
\end{eqnarray*}
we have
\begin{eqnarray*}
\nabla_{\Sigma_{x|u}}\left[
\tilde{I}_k(\Sigma_{x|u}^*) - \mu \tilde{I}_p(\Sigma_{x|u}^*) \right] = 0.
\end{eqnarray*}
By the chain rule for the derivative, we have
\begin{eqnarray*}
\lefteqn{
 \nabla_{\Sigma_{x|uz}} \left[
 \tilde{I}_k(\phi^{-1}(\Sigma_{x|uz}^*)) - \mu
  \tilde{I}_p(\phi^{-1}(\Sigma_{x|uz}^*)) \right] } \\
&=& \nabla_{\Sigma_{x|uz}} \phi^{-1}(\Sigma_{x|uz}^*) \cdot
 \nabla_{\Sigma_{x|u}} \left[
\tilde{I}_k(\Sigma_{x|u}^*) - 
\mu \tilde{I}_p(\Sigma_{x|u}^*) \right] \\
&=& 0.
\end{eqnarray*}
Thus, from Eq.~(\ref{eq:variable-changed}), we have
\begin{eqnarray}
\label{eq:kkt-4}
(\Sigma_{x|uz}^*)^{-1} = \gamma
 (\Sigma_{x|uz}^* + \Sigma_{N_3})^{-1}.
\end{eqnarray}

\noindent{\em Step 3:}
By noting that $(X,\tilde{Y},Z)$ is degraded,
from Corollary \ref{coro:cn-bound}, 
for any $(R_p,R_k) \in {\cal R}(X,\tilde{Y},Z)$ we have
\begin{eqnarray}
\lefteqn{
 R_k - \mu R_p } \nonumber \\
&\le& I(U;\tilde{Y}|Z) - \mu I(U;X|\tilde{Y}) \nonumber \\
&=& [(1+\mu) h(\tilde{Y}|Z) - \mu h(X|Z)]
 - \frac{1+\mu}{2}\log|K_{\tilde{y}x} K_{\tilde{y}x}^T| \nonumber \\
&& + \mu \left[ h(X|U,Z) - \gamma h(X + N_3|U,Z) \right],
\label{eq:upper-1}
\end{eqnarray}
where $U$ is not necessarily Gaussian.
By using the conditional version of EPI \cite{bergmans:74},
we have
\begin{eqnarray}
\lefteqn{
 h(X|U,Z) - \gamma h(X+ N_3|U,Z) } 
\label{eq:standard-extremal} \\
&\le& h(X|U,Z) \nonumber \\
&& -\frac{\gamma m}{2} \log\left(
  \exp\left[ \frac{2}{m} h(X|U,Z) \right]
  + \exp\left[ \frac{2}{m} h(N_3) \right] \right) \nonumber \\
&\le& f\left( h(N_3) - \frac{m}{2} \log(\gamma -1); h(N_3) \right),
\label{eq:epi-bound}
\end{eqnarray}
where we set
\begin{eqnarray*}
f(t;a) := t - \frac{\gamma m}{2} \log \left(
 \exp\left[ \frac{2}{m} t \right] + \exp\left[ \frac{2}{m} a \right] \right).
\end{eqnarray*}
Note that the function $f(t;a)$ is concave function of $t$ and
takes the maximum at $t = a - \frac{m}{2}\log(\gamma - 1)$ \cite{liu:07}.
From Eq.~(\ref{eq:kkt-4}), we have
\begin{eqnarray*}
(\gamma - 1)^{-1} \Sigma_{N_3}
 = \Sigma_{x|uz}^*,
\end{eqnarray*} 
which implies
\begin{eqnarray*}
h(N_3) - \frac{m}{2}\log(\gamma-1)
&=& \frac{1}{2}\log (2 \pi e)^m (\gamma -1)^{-m}|\Sigma_{N_3}| \\
&=& \frac{1}{2} \log(2 \pi e)^m |\Sigma_{x|uz}^*|.
\end{eqnarray*}
Furthermore, since $\Sigma_{x|uz}^*$ and 
$\Sigma_{N_3}$ are proportional to each other,
we have
\begin{eqnarray*}
|\Sigma_{x|uz}^*|^{1/m} + |\Sigma_{N_3}|^{1/m}
 = |\Sigma_{x|uz}^* + \Sigma_{N_3}|^{1/m}.
\end{eqnarray*}
Thus, from Eqs.~(\ref{eq:upper-1}) and (\ref{eq:epi-bound}), we have
\begin{eqnarray*}
\lefteqn{
 R_k - \mu R_p } \\
&\le& [(1+\mu) h(\tilde{Y}|Z) - \mu h(X|Z)]
 - \frac{1+\mu}{2}\log|K_{\tilde{y}x} K_{\tilde{y}x}^T| \\
&& +\mu \left[
 \frac{1}{2} \log(2\pi e)^m |\Sigma_{x|uz}^*|
  - \frac{\gamma}{2} \log (2 \pi e)^m |\Sigma_{x|uz}^* + \Sigma_{N_3} | \right] \\
&=& \tilde{I}_k(\Sigma_{x|u}^*) - \mu \tilde{I}_p(\Sigma_{x|u}^*).
\end{eqnarray*}
\qed

%%%%%%% Remark %%%%%%%%%%%%%%%%
\begin{remark}
\label{remark:extremal}
One of the difficulties in the above proof
is that, after Step 1,
 we have to show the extremal
inequality of the form
\begin{eqnarray}
&& \hspace{-10mm} \mu h(X|U) + h(X + W_z|U)
- (1+\mu) h(X + \tilde{W}_y|U)  \nonumber \\
&\le& \frac{\mu}{2} \log |\Sigma_{x|u}^*|
  + \frac{1}{2} \log |\Sigma_{x|u}^* + \Sigma_{W_z} | \nonumber \\
&& ~~~~~- \frac{(1+\mu)}{2} \log |\Sigma_{x|u}^* + \Sigma_{\tilde{W}_y} |.
\label{eq:key-extremal-inequality}
\end{eqnarray}
This type of extremal inequality has appeared in
\cite[Corollary 2]{liu:10b} (scalar version has appeared in \cite[Lemma 1]{chen:09}).
In \cite{liu:10}, the extremal inequality was proved by using
 a vector generalization of Costa's entropy power 
inequality \cite{costa:85}. On the otherhand,
we showed Eq.~(\ref{eq:key-extremal-inequality}) by using
the change of variable in Step 2 and by reducing
to more tractable
form (Eq.~(\ref{eq:standard-extremal})), 
which has appeared in the literature \cite{liu:07}.
By this reduction, we only need the standard EPI in our proof
instead of Costa's type EPI, and our proof seems more
elementary.
\end{remark} 

%%%%%%%%%% Proof of Theorem 1%%%%%%%%%%
\section{Proof of Theorem \ref{theorem:general-case}}
\label{sec:proof-general}

In this section, we show Theorem \ref{theorem:general-case}
by using Theorem \ref{theorem:aligned-case}.
We follow a similar approach as in \cite[Section 4]{ly:09}.
Since the direct part can be proved by taking
a Gaussian auxiliary random variable $U$ in 
Proposition \ref{proposition:direct-part} 
(see Section \ref{subsec:direct-aligned}),
we concentrate on the converse part.
Without loss of generality, we can assume that
the matrices $B$ and $E$ are square (but not necessarily invertible).
If that is not the case, we can apply singular value decomposition (SVD)
to show equivalent sources $(X^\prime,Y^\prime, Z^\prime)$ on
$\mathbb{R}^{m_x} \times \mathbb{R}^{m_x} \times \mathbb{R}^{m_x}$ such
that ${\cal R}(X^\prime,Y^\prime,Z^\prime) = {\cal R}(X,Y,Z)$
in a similar manner as \cite[Section 5-B]{weingarten:06}.

By using SVD, we can write the matrices as 
\begin{eqnarray*}
B &=& U_y \Lambda_y V_y, \\
E &=& U_z \Lambda_z V_z,
\end{eqnarray*}
where $U_y,V_y, U_z$ and $V_z$ are $m_x \times m_x$ orthogonal matrices,
and $\Lambda_y$ and $\Lambda_z$ are diagonal matrices. Let 
\begin{eqnarray*}
\bar{B} &=& U_y (\Lambda_y + \alpha I) V_y, \\
\bar{E} &=& U_y (\Lambda_z + \alpha I) V_z
\end{eqnarray*}
for some $\alpha > 0$. Then, let
\begin{eqnarray*}
\bar{Y} &=& \bar{B} X + W_y, \\
\bar{Z} &=& \bar{E} X + W_z.
\end{eqnarray*}
Since $\bar{B}$ and $\bar{E}$ are invertible, 
Theorem \ref{theorem:aligned-case} implies
\begin{eqnarray}
\label{eq:pertarbated-equality}
{\cal R}(X, \bar{Y}, \bar{Z}) = {\cal R}_G(X,\bar{Y},\bar{Z}).
\end{eqnarray}
In the following, we will show the following lemma.

\begin{lemma}
\label{lemma:pertarbated}
We have
\begin{eqnarray*}
{\cal R}(X,Y,Z) \subset {\cal R}(X,\bar{Y},\bar{Z}) 
 + {\cal O}(X, \bar{Y}, \bar{Z}),
\end{eqnarray*}
where 
\begin{eqnarray*}
{\cal O}(X,\bar{Y},\bar{Z})
 &=& \left\{ (0, R_k) :
 0 \le R_k \le \frac{1}{2} \log |\bar{E} \Sigma_x \bar{E}^T + I| \right. 
 \\
 && \left. ~~- \frac{1}{2} \log |E \Sigma_x E^T + I| \right\}.
\end{eqnarray*}
\end{lemma}

By letting $\alpha \to 0$, ${\cal R}_G(X, \bar{X}, \bar{Z})$
converges to ${\cal R}_G(X,Y,Z)$ and ${\cal O}(X, \bar{Y},\bar{Z})$
converges to $\{(0,0) \}$.
Thus, Eq.~(\ref{eq:pertarbated-equality}) and Lemma \ref{lemma:pertarbated} 
imply ${\cal R}(X,Y,Z) \subset {\cal R}_G(X,Y,Z)$.

\noindent{\em Proof of Lemma \ref{lemma:pertarbated}}

Let 
\begin{eqnarray*}
C_y &=& U_y \Lambda_y (\Lambda_y + \alpha I)^{-1} V_y, \\
C_z &=& U_z \Lambda_z (\Lambda_z + \alpha I)^{-1} V_z.
\end{eqnarray*}
Then, we have $C_y C_y^T \prec I$ and $C_z C_z^T \prec I$.
Thus, we can write 
\begin{eqnarray*}
Y &=& C_y \bar{Y} + W_y^\prime, \\
Z &=& C_z \bar{Z} + W_z^\prime
\end{eqnarray*}
for $W_y^\prime \sim {\cal N}(0, I - C_y C_y^T)$
and $W_z^\prime \sim {\cal N}(0, I- C_z C_z^T)$, i.e., we have
\begin{eqnarray}
\label{eq:markov-ralations-1}
&& X \markov \bar{Y} \markov Y, \\
\label{eq:markov-ralations-2}
&& X \markov \bar{Z} \markov Z. 
\end{eqnarray}

From Proposition \ref{prop:cn-bound}, for any $(R_p, R_k) \in {\cal R}(X,Y,Z)$,
there exist $(U,V)$ satisfying 
\begin{eqnarray*}
R_p &\ge& I(U;X) - I(U; Y), \\
R_k &\le& I(U;Y|V) - I(U; Z|V),
\end{eqnarray*}
and $(U,V) \markov X \markov (Y,Z)$.
Let 
\begin{eqnarray*}
\bar{R}_p &=& I(U; X) - I(U; \bar{Y}), \\
\bar{R}_k &=& I(U; \bar{Y}|V) - I(U; \bar{Z}|V).
\end{eqnarray*}
Then, we have
\begin{eqnarray*}
R_p - \bar{R}_p &\ge&
 I(U; X) - I(U;Y) - [ I(U; X) - I(U; \bar{Y}) ] \\
 &=& I(U; \bar{Y}) - I(U; Y) \\
 &\ge & 0,
\end{eqnarray*}
where the second inequality follows from
Eq.~(\ref{eq:markov-ralations-1}).
On the other hand, we have
\begin{eqnarray*}
R_k - \bar{R}_k &\le&
 I(U; Y|V) - I(U;Z|V) \\
 && - [ I(U; \bar{Y}|V) - I(U; \bar{Z}|V) ] \\
&=& I(U; \bar{Z}|V) - I(U;Z|V) \\
 && - [ I(U;\bar{Y}|V) - I(U;Y|V) ] \\
&\le& I(U; \bar{Z}|V) - I(U; Z|V) \\
&=& I(U,V; \bar{Z}) - I(U, V; Z) \\
 && - [ I(V; \bar{Z}) - I(V; Z) ] \\
&\le& I(U,V; \bar{Z}) - I(U,V; Z) \\
&=& I(X; \bar{Z}) - I(X; Z) \\
 && - [ I(X; \bar{Z}|U,V) - I(X; Z|U,V) ] \\
&\le& I(X; \bar{Z}) - I(X; Z) \\
&=& \frac{1}{2} \log |\bar{E} \Sigma_x \bar{E}^T + I|
  - \frac{1}{2} \log |E \Sigma_x E^T + I|,
\end{eqnarray*}
where the second, third, and forth inequalities follow
from the Markov relations in
Eqs.~(\ref{eq:markov-ralations-1}) and (\ref{eq:markov-ralations-2}).
\qed

\section{Conclusion}
\label{sec:conclusion}
%The conclusion goes here.

In this paper, we investigated the secret key
agreement from vector Gaussian sources by rate
limited public communication. We characterized
the optimal trade-off between the key rate
and the public communication rate as a
(covariance) matrix optimization problem.
Investigating an efficient method to solve
the optimization problem is a future research agenda.

%%%%%%%%%% Appendix %%%%%%%%%%%%%%%%%%%%
\appendices
\section{Derivation of the KKT condition}
\label{appendix:kkt}

We first rewrite the optimization problem
in Eq.~(\ref{eq:opt1}) as a standard form
\begin{eqnarray}
\mbox{minimize} && - I_k(\Sigma_{x|u}) \nonumber \\
\mbox{subject to} && I_p(\Sigma_{x|u}) - R_p^o \le 0 
 \label{eq:opt2} \\
&& 0 \prec \Sigma_{x|u} \preceq \Sigma_x.
\nonumber
\end{eqnarray}
Let $\Sigma_{x|u}^*$ be an optimal solution
for this problem, which is also an optimal solution
of Eq.~(\ref{eq:opt1}).
Then, we have $\Sigma_{x|u}^* \succ 0$ because
of the constraint $I_p(\Sigma_{x|u}^*) - R_p^o \le 0$.
Thus, there exists a positive definite
matrix $L$ satisfying $L \prec \Sigma_{x|u}^*$.

Let us consider another optimization problem
\begin{eqnarray}
\mbox{minimize} && - I_k(\Sigma_{x|u}) \nonumber \\
\mbox{subject to} && I_p(\Sigma_{x|u}) - R_p^o \le 0 
 \label{eq:opt3} \\
&& L \preceq \Sigma_{x|u} \preceq \Sigma_x.
\nonumber
\end{eqnarray}
Obviously, $\Sigma_{x|u}^*$ is also an optimal solution
for the problem in Eq.~(\ref{eq:opt3}), and
the optimal values for Eqs.~(\ref{eq:opt2})
and (\ref{eq:opt3}) are the same.
Although the optimization problem in
Eq.~(\ref{eq:opt3}) is not convex, there exist
Lagrange multipliers $M_1 \succeq 0$,
$M_2 \succeq 0$, and $\mu \ge 0$ satisfying
\begin{eqnarray}
&& \hspace{-40mm} - \left(
- \nabla_{\Sigma_{x|u}} I_k(\Sigma_{x|u}^*)
 + \mu \nabla_{\Sigma_{x|u}}(I_p(\Sigma_{x|u}^*) - R_p^o)
\right) \nonumber \\
&=& M_2 - M_1, \\
M_1(\Sigma_{x|u}^* - L) &=& 0, \label{kkt:app-2} \\
M_2(\Sigma_x - \Sigma_{x|u}^*)) &=& 0, \\
\mu (R_p - I_p(\Sigma_{x|u}^*)) &=& 0
\end{eqnarray}
if the set of constraint qualifications
(CQs) shown below are satisfied
(see \cite[Appendix 4]{weingarten:06} for
the detail).
Since $\Sigma_{x|u}^* \succ L$, 
Eq.~(\ref{kkt:app-2}) implies $M_1 = 0$.
Thus, by noting the relation
\begin{eqnarray}
\lefteqn{ I_k(\Sigma_{x|u}) - \mu I_p(\Sigma_{x|u}) } \nonumber \\
&=& \frac{\mu}{2}\log (2 \pi e)^m |\Sigma_{x|u}| 
 + \frac{1}{2} \log (2 \pi e)^m |\Sigma_{x|u} + \Sigma_{W_z}| \nonumber \\
 &&~~~~~~ - \frac{(1+\mu)}{2} \log (2 \pi e)^m |\Sigma_{x|u} + \Sigma_{W_y}| 
\nonumber\\
&& + [(1+\mu)h(Y) - h(Z) - \mu h(X)],
\label{eq:relation}
\end{eqnarray}
and by setting $M = 2M_2$, we have
the KKT conditions in Eqs.~(\ref{eq:kkt-1})--(\ref{eq:kkt-3}).

The CQs shown in \cite[Appendix 4]{weingarten:06},
which is an interpretation of 
\cite[CQ5a of Section 5.4]{bertsekas:03} are
the following:
There exists a matrix $A$ satisfying
\begin{enumerate}
\item \label{cq-1}
For any $\bol{u} \neq 0$ in the null space
of $\Sigma_{x|u}^* - L$, we have $\bol{u}^T A \bol{u} > 0$.

\item \label{cq-2}
For any $\bol{v} \neq 0$ in the null space 
of $\Sigma_x - \Sigma_{x|u}^*$, we have
$\bol{v}^T A \bol{v} < 0$.

\item \label{cq-3}
\begin{eqnarray*}
\rom{Tr}\left[
\nabla_{\Sigma_{x|u}} (I_p(\Sigma_{x|u}^*) - R_p^o)
 A^T
\right] > 0.
\end{eqnarray*}
\end{enumerate}

To check whether the above CQs are satisfied,
we suggest $A$ given by
\begin{eqnarray*}
A = \alpha (L - \Sigma_{x|u}^*) + (\Sigma_x - \Sigma_{x|u}^*)
\end{eqnarray*}
for $\alpha > 0$.
First we check (\ref{cq-1}). For any
$\bol{u} \neq 0$ in the null space 
of $\Sigma_{x|u}^* - L$, we have
\begin{eqnarray*}
\bol{u}^T A \bol{u} = \bol{u}^T (\Sigma_x - \Sigma_{x|u}^*) \bol{u}.
\end{eqnarray*}
Suppose that $\bol{u}^T (\Sigma_x - \Sigma_{x|u}^*) \bol{u} = 0$.
Then we have
\begin{eqnarray*}
0 &=& \bol{u}^T \left(
(\Sigma_{x|u}^* - L) + (\Sigma_x - \Sigma_{x|u}^*)
\right) \bol{u} \\
&=& \bol{u}^T (\Sigma_x - L) \bol{u},
\end{eqnarray*}
which is a contradiction because $\Sigma_x \succ L$.
Thus the condition (\ref{cq-1}) is satisfied.

Next, we check (\ref{cq-2}). For any
$\bol{v} \neq 0$ in the null space of
$\Sigma_x - \Sigma_{x|u}^*$, we have
\begin{eqnarray*}
\bol{v}^T A \bol{v} 
= \bol{v}^T (L-\Sigma_{x|u}^*) \bol{v} < 0
\end{eqnarray*}
because $L \prec \Sigma_{x|u}^*$.

Finally, we check (\ref{cq-3}).
By noting
\begin{eqnarray*}
\nabla_{\Sigma_{x|u}} I_p(\Sigma_{x|u})
 = \frac{1}{2} (\Sigma_{x|u} + \Sigma_{W_y})^{-1}
 - \frac{1}{2} \Sigma_{x|u}^{-1} \prec 0
\end{eqnarray*}
for any $\Sigma_{x|u} \succ 0$, we have
\begin{eqnarray*}
\lefteqn{
\rom{Tr}\left[
\nabla_{\Sigma_{x|u}}(I_p(\Sigma_{x|u}^*) - R_p^o)
A \right]
} \\
&=& \frac{\alpha}{2} \rom{Tr} \left[
\left\{
(\Sigma_{x|u}^* + \Sigma_{W_y})^{-1}
 - (\Sigma_{x|u}^*)^{-1}
\right\} (L - \Sigma_{x|u}^*)
\right] \\
&& \hspace{-3mm} + 
\frac{1}{2} \rom{Tr} \left[
\left\{
(\Sigma_{x|u}^* + \Sigma_{W_y})^{-1}
 - (\Sigma_{x|u}^*)^{-1}
\right\} (\Sigma_x - \Sigma_{x|u}^*)
\right].
\end{eqnarray*}
Since $L - \Sigma_{x|u}^* \prec 0$, by taking
$\alpha > 0$ to be sufficiently large, 
the condition (\ref{cq-3}) is satisfied.

\begin{remark}
We need to introduce the optimization problem
in Eq.~(\ref{eq:opt3}) because the arguments
in \cite[Appendix 4]{weingarten:06} is guaranteed
only under the condition such that
the range of the variable $\Sigma_{x|u}$
is a closed set.
\end{remark}

%%%%%%%%%%%%%%%%%%%%%%%%%%%%%%%%%%%%%%%%%
\section{Proof of Eqs.~(\ref{eq:enhanced-1}) and (\ref{eq:enhanced-2})}
\label{appendix:enhanced-1-2}

By noting $M \succeq 0$, we have
\begin{eqnarray*}
(\Sigma_{x|u}^* + \Sigma_{\tilde{W}_y})^{-1}
 &=& (\Sigma_{x|u}^* + \Sigma_{W_y})^{-1} +  M \\
&\succeq& (\Sigma_{x|u}^* + \Sigma_{W_y})^{-1}.
\end{eqnarray*}
Thus we have
\begin{eqnarray*}
\Sigma_{\tilde{W}_y} \preceq \Sigma_{W_y}.
\end{eqnarray*}

Since $\Sigma_{W_z} \succ 0$, by substituting
Eq.~(\ref{eq:enhancement}) into Eq.~(\ref{eq:kkt-1}),
we have
\begin{eqnarray}
\lefteqn{
(\Sigma_{x|u}^* + \Sigma_{\tilde{W}_y})^{-1}
} \nonumber \\
&=& \frac{\mu}{1+\mu} (\Sigma_{x|u}^*)^{-1}
 + \frac{1}{1+\mu}(\Sigma_{x|u}^* + \Sigma_{W_z})^{-1} 
\label{eq:proof-enhanced-1} \\
&\prec& (\Sigma_{x|u}^*)^{-1}
\nonumber
\end{eqnarray}
when $\mu > 0$.
Thus, we have
\begin{eqnarray*}
\Sigma_{\tilde{W}_y} \succ 0.
\end{eqnarray*}
Note that $\Sigma_{\tilde{W}_y} = \Sigma_{W_z} \prec 0$
when $\mu = 0$.

From Eq.~(\ref{eq:proof-enhanced-1}), we have
\begin{eqnarray*}
(\Sigma_{x|u}^* + \Sigma_{\tilde{W}_y})^{-1}
 \succeq (\Sigma_{x|u}^* + \Sigma_{W_z})^{-1},
\end{eqnarray*}
where the strict inequality holds for
$\mu > 0$. Thus we have
\begin{eqnarray*}
\Sigma_{\tilde{W}_y} \preceq \Sigma_{W_z}
\end{eqnarray*}
and especially
\begin{eqnarray}
\label{eq:strict-enhanced}
\Sigma_{\tilde{W}_y} \prec \Sigma_{W_z}
\end{eqnarray}
for $\mu > 0$.
\qed

%%%%%%% Appendix B %%%%%%%%%%%%%%%%%
\section{Proofs of Eq.~(\ref{eq:reduction})}
\label{appendix:preservation}

Eq.~(\ref{eq:reduction}) can be derived by
the following sequence of equalities:
\begin{eqnarray}
\lefteqn{
(\Sigma_x + \Sigma_{\tilde{W}_y})
 (\Sigma_{x|u}^* + \Sigma_{\tilde{W}_y})^{-1} } \nonumber \\
&=& \left[ (\Sigma_x - \Sigma_{x|u}^*) +
 (\Sigma_{x|u}^* + \Sigma_{\tilde{W}_y}) \right]
 (\Sigma_{x|u}^* + \Sigma_{\tilde{W}_y})^{-1} \nonumber \\
&=& (\Sigma_x - \Sigma_{x|u}^*)
  (\Sigma_{x|u}^* + \Sigma_{\tilde{W}_y})^{-1}
  + I \nonumber \\
&=& (\Sigma_x - \Sigma_{x|u}^*)
 \left[ (\Sigma_{x|u}^* + \Sigma_{W_y})^{-1} + M \right]
  + I \label{eq:proof-preservation-1} \\
&=& (\Sigma_x - \Sigma_{x|u}^*)
 (\Sigma_{x|u}^* + \Sigma_{W_y})^{-1} + I 
 \label{eq:proof-preservation-2} \\
&=& \left[ (\Sigma_x - \Sigma_{x|u}^*)
 + (\Sigma_{x|u}^* + \Sigma_{W_y}) \right]
 (\Sigma_{x|u}^* + \Sigma_{W_y})^{-1} \nonumber \\
&=& (\Sigma_x + \Sigma_{W_y}) 
  (\Sigma_{x|u}^* + \Sigma_{W_y})^{-1},
\end{eqnarray}
where Eq.~(\ref{eq:proof-preservation-1})
follows from Eq.~(\ref{eq:enhancement}) and
Eq.~(\ref{eq:proof-preservation-2}) follows
from Eq.~(\ref{eq:kkt-2}).
%%%%%%% Appendix C %%%%%%%%%%%%%%%%%%%%%%%
\section{Proof of Eq.~(\ref{eq:invertibility})}
\label{appendix:invertibility}

From Eqs.~(\ref{eq:enhanced-y}),
(\ref{eq:aligned-z}) and (\ref{eq:strict-enhanced}),
we can write
\begin{eqnarray}
Z = X + \tilde{W}_y + W^\prime,
\end{eqnarray}
where $W^\prime \sim {\cal N}(0,\Sigma_{\tilde{W}_y} - \Sigma_{W_z})$.
Thus, we have
\begin{eqnarray*}
\Sigma_{\tilde{y}z} &=& \Sigma_{\tilde{y}}, \\
\Sigma_{zx} &=& \Sigma_{xz}
 = \Sigma_x.
\end{eqnarray*}
Furthermore, we have
\begin{eqnarray}
\label{eq:proof-invertiblity-0}
\Sigma_{\tilde{y}} \prec \Sigma_z.
\end{eqnarray}

From the block inversion of the matrix
(e.g.~see \cite[Appendix 5.5]{boyd-book:04}) and
Eq.~(\ref{eq:coefficients}), we have
\begin{eqnarray}
K_{\tilde{y}x}
 &=& 
 \left[ \begin{array}{cc}
\Sigma_{\tilde{y}} & \Sigma_x 
 \end{array}\right]
\left[
\begin{array}{c}
- \Sigma_z^{-1} \Sigma_x S^{-1} \\
 S^{-1}
\end{array}\right] 
 \nonumber \\
 &=& (I - \Sigma_{\tilde{y}} \Sigma_z^{-1}) \Sigma_x S^{-1},
\label{eq:proof-invertibility-1}
\end{eqnarray}
where
\begin{eqnarray*}
S = \Sigma_x - \Sigma_x \Sigma_z^{-1} \Sigma_x
\end{eqnarray*}
is the Schur complement.

From Eq.~(\ref{eq:proof-invertiblity-0}),
we have
\begin{eqnarray*}
I - \Sigma_{\tilde{y}}^{\frac{1}{2}}
 \Sigma_z^{-1} \Sigma_{\tilde{y}}^{\frac{1}{2}}
 \succ I - \Sigma_{\tilde{y}}^{\frac{1}{2}}
 \Sigma_{\tilde{y}}^{-1} \Sigma_{\tilde{y}}^{\frac{1}{2}}
 = 0.
\end{eqnarray*}
Thus we have
\begin{eqnarray}
\label{eq:proof-invertibility-2}
| I - \Sigma_{\tilde{y}} \Sigma_z^{-1}|
 = | I - \Sigma_{\tilde{y}}^{\frac{1}{2}}
 \Sigma_z^{-1} \Sigma_{\tilde{y}}^{\frac{1}{2}}|
\neq 0.
\end{eqnarray}
By combining Eqs.~(\ref{eq:proof-invertibility-1}) 
and (\ref{eq:proof-invertibility-2}),
we have Eq.~(\ref{eq:invertibility}).
\qed

%%%%%%%%%%% Ack %%%%%%%%%%%%%%%%%%%%%%%%%%%%%

\section*{Acknowledgment}

The first author would like to thank 
Prof.~Ryutaroh Matsumoto
for valuable discussions and comments.
The authors also would like to thank
Prof.~Jun Chen for informing the literatures
\cite{liu:10b,chen:09}.
This research is partly supported by 
Grant-in-Aid for Young Scientists (Start-up):
KAKENHI 21860064.

%\bibliographystyle{../09-04-17-bibtex/IEEEtran}
%\bibliography{../09-04-17-bibtex/reference.bib}

\begin{thebibliography}{10}
\providecommand{\url}[1]{#1}
\csname url@samestyle\endcsname
\providecommand{\newblock}{\relax}
\providecommand{\bibinfo}[2]{#2}
\providecommand{\BIBentrySTDinterwordspacing}{\spaceskip=0pt\relax}
\providecommand{\BIBentryALTinterwordstretchfactor}{4}
\providecommand{\BIBentryALTinterwordspacing}{\spaceskip=\fontdimen2\font plus
\BIBentryALTinterwordstretchfactor\fontdimen3\font minus
  \fontdimen4\font\relax}
\providecommand{\BIBforeignlanguage}[2]{{%
\expandafter\ifx\csname l@#1\endcsname\relax
\typeout{** WARNING: IEEEtran.bst: No hyphenation pattern has been}%
\typeout{** loaded for the language `#1'. Using the pattern for}%
\typeout{** the default language instead.}%
\else
\language=\csname l@#1\endcsname
\fi
#2}}
\providecommand{\BIBdecl}{\relax}
\BIBdecl

\bibitem{ahlswede:93}
R.~Ahlswede and I.~Csisz\'ar, ``Common randomness in information theory and
  cryptography--part {I}: Secret sharing,'' \emph{IEEE Trans. Inform. Theory},
  vol.~39, no.~4, pp. 1121--1132, July 1993.

\bibitem{csiszar:00}
I.~Csisz\'ar and P.~Narayan, ``Common randomness and secret key generation with
  a helper,'' \emph{IEEE Trans. Inform. Theory}, vol.~46, no.~2, pp. 344--366,
  March 2000.

\bibitem{csiszar:04}
------, ``Secrecy capacities for multiple terminals,'' \emph{IEEE Trans.
  Inform. Theory}, vol.~50, no.~12, pp. 3047--3061, December 2004.

\bibitem{maurer:93}
U.~Maurer, ``Secret key agreement by public discussion from common
  information,'' \emph{IEEE Trans. Inform. Theory}, vol.~39, no.~3, pp.
  733--742, May 1993.

\bibitem{wyner:75}
A.~D. Wyner, ``The wire-tap channel,'' \emph{Bell Syst.~Tech.~J.}, vol.~54,
  no.~8, pp. 1355--1387, 1975.

\bibitem{csiszar:78}
I.~Csisz\'ar and J.~K\"orner, ``Broadcast channels with confidential
  messages,'' \emph{IEEE Trans. Inform. Theory}, vol.~24, no.~3, pp. 339--348,
  May 1979.

\bibitem{liang:08}
Y.~Liang, H.~V. Poor, and S.~Shamai, ``Secure communication over fading
  channels,'' \emph{IEEE Trans. Inform. Theory}, vol.~54, no.~6, pp.
  2470--2492, June 2008.

\bibitem{liu:09}
T.~Liu and S.~Shamai, ``A note on the secrecy capacity of the multiple-antenna
  wiretap channel,'' \emph{IEEE Trans. Inform. Theory}, vol.~55, no.~6, pp.
  2547--2553, June 2009.

\bibitem{bustin:09}
R.~Bustin, R.~Liu, H.~V. Poor, and S.~Shamai, ``An {MMSE} approach to the
  secrecy capacity of the {MIMO} {G}aussian wiretap channel,'' \emph{EURASIP
  Journal on Wireless Communication and Networking}, 2009.

\bibitem{ly:09}
H.~D. Ly, T.~Liu, and Y.~Liang, ``Multiple-input multiple-output gaussian
  broadcast channels with common and confidential messages,'' 2009,
  arXiv:0907.2599.

\bibitem{liu:09b}
R.~Liu, T.~Liu, V.~Poor, and S.~{Shamai (Shitz)}, ``Multiple-input
  multiple-output gaussian broadcast channels with confidential messages,''
  2009, arXiv:0903.3786.

\bibitem{ekrem:09}
E.~Ekrem and S.~Ulukus, ``The secrecy capacity region of the gaussian {MIMO}
  multi-receiver wiretap channel,'' 2009, arXiv:0903.3096v1.

\bibitem{ekrem:10}
------, ``Gaussian {MIMO} broadcast channels with common and confidential
  messages,'' in \emph{Proc. IEEE Int. Symp. Inf. Theory 2010}, Austin, U.S.A.,
  2010, pp. 2583--2587.

\bibitem{liu:10}
R.~Liu, T.~Liu, H.~V. Poor, and S.~{Shamai (Shitz)}, ``{MIMO} gaussian
  broadcast channels with confidential and common messages,'' in \emph{Proc.
  IEEE Int. Symp. Inf. Theory 2010}, Austin, U.S.A., 2010, pp. 2578--2582.

\bibitem{khisti:10}
A.~Khisti and G.~W. Wornell, ``Secure transmission with multiple antennas
  {I}:the misome wiretap channel,'' \emph{IEEE Trans. Inform. Theory}, vol.~56,
  no.~7, pp. 3088--3104, July 2010.

\bibitem{bloch:08}
M.~Bloch, J.~Barros, M.~R.~D. Rodrigues, and S.~W. McLaughlin, ``Wireless
  information-theoretic security,'' \emph{IEEE Trans. Inform. Theory}, vol.~54,
  no.~6, pp. 2515--2534, June 2008.

\bibitem{grosshans:03}
F.~Grosshans, G.~V. Assche, J.~Wenger, R.~Brouri, N.~J. Cerf, and P.~Grangier,
  ``Quantum key distribution using {G}aussian-modulated coherent states,''
  \emph{Nature}, vol. 421, pp. 238--241, January 2003.

\bibitem{assche:04}
G.~V. Assche, J.~Cardinal, and N.~J. Cerf, ``Reconciliation of a
  quantum-distributed {G}aussian key,'' \emph{IEEE Trans. Inform. Theory},
  vol.~50, no.~2, pp. 394--400, February 2004.

\bibitem{assche:book}
G.~V. Assche, \emph{Quantum Cryptography and Secret-Key Distillation}.\hskip
  1em plus 0.5em minus 0.4em\relax Cambridge Univ.~Press, 2006.

\bibitem{watanabe:09c}
S.~Watanabe and Y.~Oohama, ``Secret key agreement from correlated {G}aussian
  sources by rate limited public communication,'' \emph{IEICE Trans.
  Fundamentals (to be published)}, vol. E93A, no.~11, 2010, arXiv:1001.3705.

\bibitem{cover}
T.~M. Cover and J.~A. Thomas, \emph{Elements of Information Theory},
  2nd~ed.\hskip 1em plus 0.5em minus 0.4em\relax John Wiley \& Sons, 2006.

\bibitem{weingarten:06}
H.~Weingarten, Y.~Steinberg, and S.~{Shamai (Shitz)}, ``The capacity region of
  the {G}aussian {MIMO} broadcast channel,'' \emph{IEEE Trans. Inform. Theory},
  vol.~52, no.~9, pp. 3936--3964, September 2006.

\bibitem{strang}
G.~Strang, \emph{Linear Algebra and Its Application}.\hskip 1em plus 0.5em
  minus 0.4em\relax Wellesley-Cambridge Press, 1998.

\bibitem{weingarten:06b}
H.~Weingarten, Y.~Steinberg, and S.~{Shamai (Shitz)}, ``On the capacity region
  of the multi-antenna broadcast channel with common messages,'' in \emph{Proc.
  IEEE Int. Symp. Inf. Theory 2006}, Seattle, WA, 2006, pp. 2195--2199.

\bibitem{weingarten:09}
H.~Weingarten, T.~Liu, S.~Shamai, Y.~Steinberg, and P.~Viswanath, ``The
  capacity region of the degraded multiple-input multiple-output compound
  broadcast channel,'' \emph{IEEE Trans. Inform. Theory}, vol.~55, no.~11, pp.
  5011--5023, November 2009.

\bibitem{bergmans:74}
P.~P. Bergmans, ``A simple converse for broadcast channels with additive white
  {G}aussian noise,'' \emph{IEEE Trans. Inform. Theory}, vol.~20, no.~2, pp.
  279--280, March 1974.

\bibitem{liu:07}
T.~Liu and P.~Viswanath, ``An extremal inequality motivated by multiterminal
  information-theoretic problems,'' \emph{IEEE Trans. Inform. Theory}, vol.~53,
  no.~5, pp. 1839--1851, May 2007.

\bibitem{liu:10b}
R.~Liu, T.~Liu, H.~V. Poor, and S.~{Shamai (Shitz)}, ``A vector generalization
  of costa's entropy-power inequality with applications,'' \emph{IEEE Trans.
  Inform. Theory}, vol.~56, no.~4, pp. 1865--1879, April 2010.

\bibitem{chen:09}
J.~Chen, ``Rate region of gaussian multiple description coding with individual
  and central distortion constraints,'' \emph{IEEE Trans. Inform. Theory},
  vol.~55, no.~9, pp. 3991--4005, September 2009.

\bibitem{costa:85}
M.~H.~M. Costa, ``A new entropy power inequality,'' \emph{IEEE Trans. Inform.
  Theory}, vol.~31, no.~6, pp. 751--760, November 1985.

\bibitem{bertsekas:03}
D.~P. Bertsekas, A.~Nedi\'c, and A.~E. Ozdaglar, \emph{Convex Analysis and
  Optimization}.\hskip 1em plus 0.5em minus 0.4em\relax Athena Scientific,
  2003.

\bibitem{boyd-book:04}
S.~Boyd and L.~Vandenberghe, \emph{Convex Optimization}.\hskip 1em plus 0.5em
  minus 0.4em\relax Cambridge University Press, 2004.

\end{thebibliography}

% Generated by IEEEtran.bst, version: 1.12 (2007/01/11)

%%%%% Author Bib %%%%%%%%%%%%%%%%%

\end{document}